\PassOptionsToPackage{unicode}{hyperref}
\PassOptionsToPackage{hyphens}{url}
\PassOptionsToPackage{dvipsnames,svgnames,x11names}{xcolor}
\documentclass[
]{article}
\usepackage{xcolor}
\usepackage{amsmath,amssymb}
\usepackage{iftex}
  \usepackage[T1]{fontenc}
  \usepackage[utf8]{inputenc}
  \usepackage{textcomp} 
\usepackage{lmodern}
\IfFileExists{upquote.sty}{\usepackage{upquote}}{}
\IfFileExists{microtype.sty}{
  \usepackage[]{microtype}
  \UseMicrotypeSet[protrusion]{basicmath} 
}{}
\makeatletter
\@ifundefined{KOMAClassName}{
  \IfFileExists{parskip.sty}{%
    \usepackage{parskip}
  }{
    \setlength{\parindent}{0pt}
    \setlength{\parskip}{6pt plus 2pt minus 1pt}}
}{
  \KOMAoptions{parskip=half}}
\makeatother
\makeatletter
\ifx\paragraph\undefined\else
  \let\oldparagraph\paragraph
  \renewcommand{\paragraph}{
    \@ifstar
      \xxxParagraphStar
      \xxxParagraphNoStar
  }
  \newcommand{\xxxParagraphStar}[1]{\oldparagraph*{#1}\mbox{}}
  \newcommand{\xxxParagraphNoStar}[1]{\oldparagraph{#1}\mbox{}}
\fi
\ifx\subparagraph\undefined\else
  \let\oldsubparagraph\subparagraph
  \renewcommand{\subparagraph}{
    \@ifstar
      \xxxSubParagraphStar
      \xxxSubParagraphNoStar
  }
  \newcommand{\xxxSubParagraphStar}[1]{\oldsubparagraph*{#1}\mbox{}}
  \newcommand{\xxxSubParagraphNoStar}[1]{\oldsubparagraph{#1}\mbox{}}
\fi
\makeatother

\usepackage{color}
\usepackage{fancyvrb}

\DefineVerbatimEnvironment{Highlighting}{Verbatim}{commandchars=\\\{\}}
\usepackage{framed}
\definecolor{shadecolor}{RGB}{241,243,245}
\newenvironment{Shaded}{\begin{snugshade}}{\end{snugshade}}

\newcommand{\AttributeTok}[1]{\textcolor[rgb]{0.40,0.45,0.13}{#1}}

\newcommand{\ConstantTok}[1]{\textcolor[rgb]{0.56,0.35,0.01}{#1}}

\newcommand{\DecValTok}[1]{\textcolor[rgb]{0.68,0.00,0.00}{#1}}

\newcommand{\FunctionTok}[1]{\textcolor[rgb]{0.28,0.35,0.67}{#1}}

\newcommand{\NormalTok}[1]{\textcolor[rgb]{0.00,0.23,0.31}{#1}}

\newcommand{\OtherTok}[1]{\textcolor[rgb]{0.00,0.23,0.31}{#1}}

\newcommand{\SpecialCharTok}[1]{\textcolor[rgb]{0.37,0.37,0.37}{#1}}

\newcommand{\StringTok}[1]{\textcolor[rgb]{0.13,0.47,0.30}{#1}}

\usepackage{longtable,booktabs,array}
\usepackage{calc} 
\usepackage{etoolbox}
\makeatletter
\patchcmd\longtable{\par}{\if@noskipsec\mbox{}\fi\par}{}{}
\makeatother
\IfFileExists{footnotehyper.sty}{\usepackage{footnotehyper}}{\usepackage{footnote}}
\makesavenoteenv{longtable}
\usepackage{graphicx}
\makeatletter
\newsavebox\pandoc@box
\newcommand*\pandocbounded[1]{
  \sbox\pandoc@box{#1}%
  \Gscale@div\@tempa{\textheight}{\dimexpr\ht\pandoc@box+\dp\pandoc@box\relax}%
  \Gscale@div\@tempb{\linewidth}{\wd\pandoc@box}%
  \ifdim\@tempb\p@<\@tempa\p@\let\@tempa\@tempb\fi
  \ifdim\@tempa\p@<\p@\scalebox{\@tempa}{\usebox\pandoc@box}%
  \else\usebox{\pandoc@box}%
  \fi%
}
\def\fps@figure{htbp}
\makeatother

\NewDocumentCommand\citeproctext{}{}

\makeatletter
 \let\@cite@ofmt\@firstofone
 \def\@biblabel#1{}
 \def\@cite#1#2{{#1\if@tempswa , #2\fi}}
\makeatother
\newlength{\cslhangindent}
\newlength{\csllabelwidth}
\newenvironment{CSLReferences}[2] 
 {\begin{list}{}{%
  \setlength{\itemindent}{0pt}
  \setlength{\leftmargin}{0pt}
  \setlength{\parsep}{0pt}
  \ifodd #1
   \setlength{\leftmargin}{\cslhangindent}
   \setlength{\itemindent}{-1\cslhangindent}
  \fi
  \setlength{\itemsep}{#2\baselineskip}}}
 {\end{list}}
\usepackage{calc}

\usepackage{amsfonts}
\usepackage{amsmath}
\usepackage{arxiv}
\usepackage{orcidlink}
\usepackage{amsmath}
\usepackage[T1]{fontenc}
\makeatletter
\@ifpackageloaded{caption}{}{\usepackage{caption}}
\AtBeginDocument{%
\ifdefined\contentsname
  \renewcommand*\contentsname{Table of contents}
\else
  \newcommand\contentsname{Table of contents}
\fi
\ifdefined\listfigurename
  \renewcommand*\listfigurename{List of Figures}
\else
  \newcommand\listfigurename{List of Figures}
\fi
\ifdefined\listtablename
  \renewcommand*\listtablename{List of Tables}
\else
  \newcommand\listtablename{List of Tables}
\fi
\ifdefined\figurename
  \renewcommand*\figurename{Figure}
\else
  \newcommand\figurename{Figure}
\fi
\ifdefined\tablename
  \renewcommand*\tablename{Table}
\else
  \newcommand\tablename{Table}
\fi
}
\@ifpackageloaded{float}{}{\usepackage{float}}
\floatstyle{ruled}
\@ifundefined{c@chapter}{\newfloat{codelisting}{h}{lop}}{\newfloat{codelisting}{h}{lop}[chapter]}
\floatname{codelisting}{Listing}

\makeatother
\makeatletter
\@ifpackageloaded{caption}{}{\usepackage{caption}}
\@ifpackageloaded{subcaption}{}{\usepackage{subcaption}}
\makeatother
\usepackage{bookmark}
\IfFileExists{xurl.sty}{\usepackage{xurl}}{} 
\hypersetup{
  pdftitle={Modelling phenology using ordered categorical generalized additive models},
  pdfauthor={David L Miller},
  pdfkeywords={ordered categorical, proportional odds logistic
model, generalized additive model, phenology},
  colorlinks=true,
  linkcolor={blue},
  filecolor={Maroon},
  citecolor={Blue},
  urlcolor={Blue},
  pdfcreator={LaTeX via pandoc}}

\newcommand{\runninghead}{A Preprint }
\renewcommand{\runninghead}{Phenology with GAMs }
\title{Modelling phenology using ordered categorical generalized
additive models}
\author{\textbf{David L
Miller}~\orcidlink{0000-0002-9640-6755}\\\\Biomathematics and Statistics
Scotland\\Invergowrie\\\\UK Centre for Ecology and
Hydrology\\Lancaster\\\href{mailto:dave.miller@bioss.ac.uk}{dave.miller@bioss.ac.uk}}
\date{}
\begin{document}
\maketitle
\begin{abstract}
One form of data collected in ecology is phenological, describing the
timing of life stages. It can be tempting to analyze such data using a
continuous distribution or to model individual transitions via
probit/logit models. Such simplifications can lead to incorrect
inference in various ways, all of which stem from ignoring the natural
structure of the data. This paper presents a flexible approach to
modelling ordered categorical data using the popular R package
\texttt{mgcv}. An example analysis of saxifrage phenology in Greenland
including useful plots, model checking and derived quantities is
included.
\end{abstract}
{\bfseries \emph Keywords}
\def\sep{\textbullet\ }
ordered categorical \sep proportional odds logistic
model \sep generalized additive model \sep 
phenology

\section{Introduction}\label{sec-intro}

Phenology describes the timing of life stages in biological organisms
(Macphie \& Phillimore 2024). Examples include plants (stages of flower
and leaf growth), insects (changes from one instar to the next), birds
(moult stages) and more. In many cases, data take the form of an ordered
categorical response (the life stage) with continuous predictors such as
date (though note there are many other forms such data could take, we
assume this structure here). These data can be awkward to deal with,
often requiring specialised models or assumptions about the underlying
process. Using either a specialized or overly-simplistic model often
means that checking and diagnostic procedures are lacking, leading to a
headache for those analysing the data. Here we present relatively simple
method, based on a popular model class, included in standard software,
where well-known diagnostics can be applied.

Ordered categorical regression has been implemented in the R package
\texttt{mgcv} since Wood \emph{et al.} (2016). This allows users to
specify a set of ordered categories and model the probability of being
in those categories using any of the machinery usually available when
using generalized additive models (GAMs) (Wood 2017). This opens-up a
large set of flexible potential models for this data type.

This paper first reviews the statistical ideas underlying these models,
before going on to describe several analyses of a data set on purple
saxifrage (\emph{Saxifraga oppositifolia}) in Greenland. We conclude by
suggesting possible applications and extensions.

\section{Statistical details}\label{stats}

The ordered categorical model implemented in \texttt{mgcv} is that of
Hastie \emph{et al.} (1989), which assumes that the categories we
observe in our data are related to an underlying (latent) continuous
response. This is based on prior work from McCullagh (1980).

Our observed stage, \(Z\), is the ordered categorization of the latent
variable \(\tilde{Z}\). \(Z\) takes values \(1 \leq k \leq K\) where
\(K\) is the number of stages. We map \(\tilde{Z}\) to \(Z\) by
determining which pair of the \emph{thresholds} \(\tilde{Z}\) lies
between: \(\theta_0 < \theta_1 < \ldots < \theta_{K-1} < \theta_K\)
(where \(\theta_0=-\infty\), \(\theta_1=-1\), and \(\theta_K=\infty\)).
That is, if \(Z=k\) then \(\theta_{k-1} < \tilde{Z} \leq \theta_{k}\).

We wish to model
\(\mathbb{P}(Z \leq k \vert \eta) = \mathbb{P}(\tilde{Z} \leq \theta_k \vert \eta)\)
where \(\eta\) is a GAM-like linear predictor (with random effects,
splines etc). Letting \(\tilde{Z}\vert \eta\) be logistically
distributed and with some re-arrangement we can model:
\begin{equation}\phantomsection\label{eq-ocat}{
\mathbb{P}(Z_i \leq k \vert \eta_i) = \text{logit}^{-1}(\theta_k - \eta_i).
}\end{equation} Our likelihood is then formed by taking the product over
\(i\) and \(k\). The use of a logistic distribution is not crucial to
our derivation (Hastie \emph{et al.} 1989) but is computationally
useful. Internally, \texttt{mgcv} estimates the \(\theta_k\) values as
hyperparameters relative to the parameters included in \(\eta_i\).

Letting
\(\gamma_k(x) = \text{logit}\left[ \mathbb{P}(Z_i \leq k \vert \eta_i)\right]\)
and taking two covariate values \(x_1\) and \(x_2\), the odds ratio
\(\gamma_k(x_1)(1 - \gamma_k(x_2)))/(1 -\gamma_k(x_1))\gamma_k(x_2)=\exp(\beta^t(x_2-x_1))\)
depends only on the covariate and not the stage (\(k\)), it is therefore
referred to as a \emph{proportional odds} model.

With this formulation we can now decompose
\(\eta=\beta_0 + \mathbf{X}\boldsymbol{\alpha} + \sum_j s_j(x_j)\) where
\(\beta_0\) is an intercept, \(\mathbf{X}\) are covariates with fixed
effects \(\boldsymbol{\alpha}\) and \(s_j\) are smooth functions of one
or more of the covariates \(x_j\). The \(s_j\)s are subject to penalties
(equivalently priors; Miller (2025)) which prevent overfitting and
impose structure on the flexible components of the model (Wood 2017).

\section{\texorpdfstring{Ordered Categorical Data Analysis in
\texttt{mgcv}}{Ordered Categorical Data Analysis in mgcv}}\label{ordered-categorical-data-analysis-in-mgcv}

The Greenland Ecosystem Monitoring Programme
(\url{https://g-e-m.dk/about-gem}) collects data measuring climate
impacts on Arctic ecosystems. As part of this programme, data were
collected weekly on purple saxifrage from late May to early September
1996-2023 (Greenland Ecosystem Monitoring 2020). Purple saxifrage are
creeping, low-growing plants that are amongst the first to flower,
continuing to do so through the summer and even into the autumn (Körner
2011; \emph{{BioBasis Manual}} 2019). Data were collected at three sites
(situated at within 50m longitudinally and 200m latitudinally of each
other, at fixed locations). Data were collected as counts of plants at
stage (``buds'', ``flowers'', ``senescent'' (flowers lost or browned)
and ``open'' (seeds exposed)), individual plants were not tracked (see
Discussion).

Output and plots are kept to a minimum here. The Supplementary Materials
include an ``extended'' version of the paper with additional content.
The document source contains all analysis code.

\subsection{A simple model for saxifrage phenology in
2010}\label{a-simple-model-for-saxifrage-phenology-in-2010}

\begin{figure}[h]

\centering{

\pandocbounded{\includegraphics[keepaspectratio]{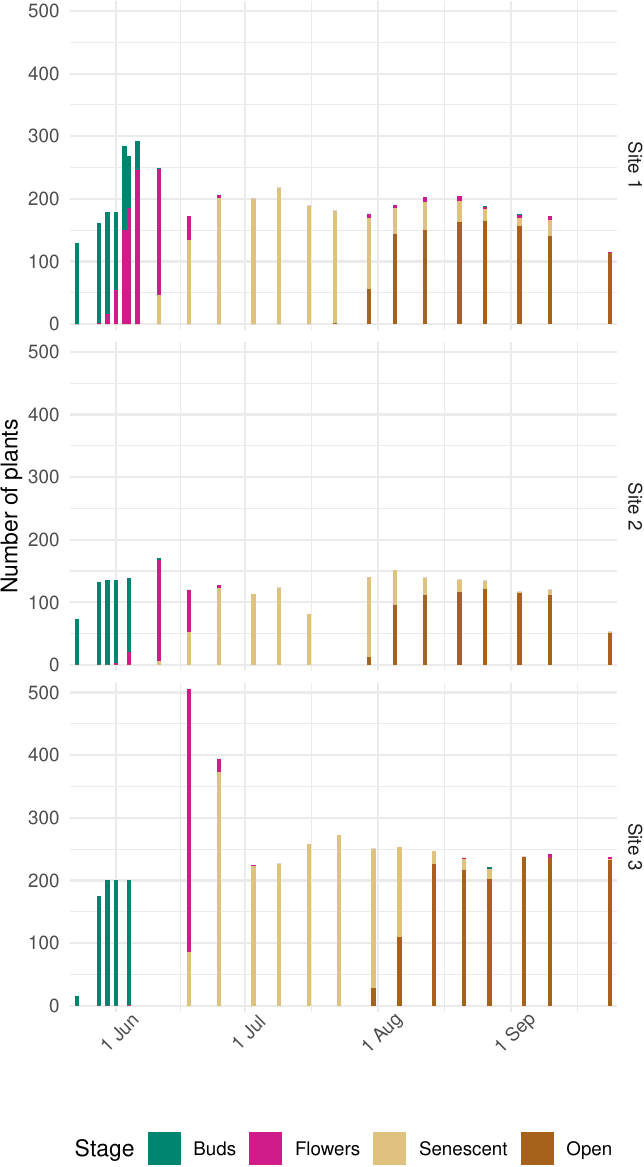}}

}

\caption{\label{fig-sax2010-raw}Raw data on purple saxifrage phenology
over three sites in Greenland during 2010. The figure is faceted by
site, which were geographically close.}

\end{figure}

Each row of the processed data corresponds to one observation of a plant
(though it is very likely, there is no guarantee that a given plant
would be counted in every visit; plants are not uniquely identified).
The variables \texttt{Date}, \texttt{Site} and \texttt{Stage} give the
date of sampling, sample site identifier and phenological stage,
respectively. \texttt{doy} and \texttt{Year} are the ordinal day of year
and year. \texttt{iStage} gives an integer version of \texttt{Stage}
required for \texttt{mgcv} to form the response variable (though note
that this does \emph{not} mean we are assuming equal distances between
the categories, it is just for convenience in the data).
Figure~\ref{fig-sax2010-raw} shows the counts per site and phenological
stage over the course of 2010 (\(n=\) 11273 non-unique plants counted in
total across all stages, sites and sampling occasions).

To begin with we construct a simple GAM that ignores \texttt{Site}:
\begin{equation}\phantomsection\label{eq-model1}{
\mathbb{P}(Z_i \leq k \vert \eta_i) = \text{logit}^{-1}(\theta_k - \eta_i)
\quad \text{where} \quad
\eta_i = \beta_0 + s(\texttt{DOY}_i),
}\end{equation} where \(\texttt{DOY}_i\) is the day of year on which the
stage \(k \in (1,4)\) was observed \(Z_i\) (where \(1,\ldots,4\)
represent the stages ``buds'', ``flowers'', ``senescent'' and ``open''
in that order). \(s\) indicates a smooth function, which we model with
thin plate regression splines (Wood 2003).

We can write this model as:

\begin{Shaded}
\begin{Highlighting}[]
\FunctionTok{library}\NormalTok{(mgcv)}

\NormalTok{b1 }\OtherTok{\textless{}{-}} \FunctionTok{gam}\NormalTok{(iStage }\SpecialCharTok{\textasciitilde{}} \FunctionTok{s}\NormalTok{(doy, }\AttributeTok{k=}\DecValTok{25}\NormalTok{),}
          \AttributeTok{data=}\NormalTok{sax\_ocat10, }\AttributeTok{family=}\FunctionTok{ocat}\NormalTok{(}\AttributeTok{R=}\DecValTok{4}\NormalTok{), }\AttributeTok{method=}\StringTok{"REML"}\NormalTok{)}
\end{Highlighting}
\end{Shaded}

The \texttt{iStage} integer variable is the response is modelled as a
smooth of \texttt{doy} and we specify the number of categories to the
\texttt{ocat} family using the \texttt{R=} argument.

We can use the \texttt{summary} function to investigate the model
results (see extended paper). The model explains 89.99\% of the
deviance. The smooth of day of year has an effective degrees of freedom
of 19.85 where the maximum complexity was 25. If the EDF is close to the
maximum complexity we would generally increase complexity, however we
are limited by the 29 unique values of \texttt{doy}. However, this model
is simple and doesn't account for other factors like \texttt{Site},
which we will address in the next section.

Calling \texttt{predict} with \texttt{type="response"} will return the
probability that saxifrage is in each stage on a given day of the year
(a matrix with one column for each stage). Setting \texttt{se=TRUE} will
give us standard errors for estimates (the function will return a
\texttt{list} with elements \texttt{\$fit} and \texttt{\$se.fit}). For
example to predict the stage probabilities for August 1st with standard
errors under \texttt{b1}:

\begin{Shaded}
\begin{Highlighting}[]
\NormalTok{pred\_aug\_1 }\OtherTok{\textless{}{-}} \FunctionTok{data.frame}\NormalTok{(}\AttributeTok{doy =} \FunctionTok{yday}\NormalTok{(}\FunctionTok{ymd}\NormalTok{(}\StringTok{"2010{-}08{-}01"}\NormalTok{)))}
\FunctionTok{predict}\NormalTok{(b1, }\AttributeTok{newdata=}\NormalTok{pred\_aug\_1, }\AttributeTok{type=}\StringTok{"response"}\NormalTok{, }\AttributeTok{se=}\ConstantTok{TRUE}\NormalTok{)}
\end{Highlighting}
\end{Shaded}

\begin{verbatim}
$fit
          [,1]         [,2]      [,3]      [,4]
1 1.390935e-06 0.0006760649 0.5505959 0.4487266

$se.fit
          [,1]        [,2]       [,3]      [,4]
1 2.669616e-07 0.000129669 0.04734797 0.0474779
\end{verbatim}

\begin{figure}

\centering{

\pandocbounded{\includegraphics[keepaspectratio]{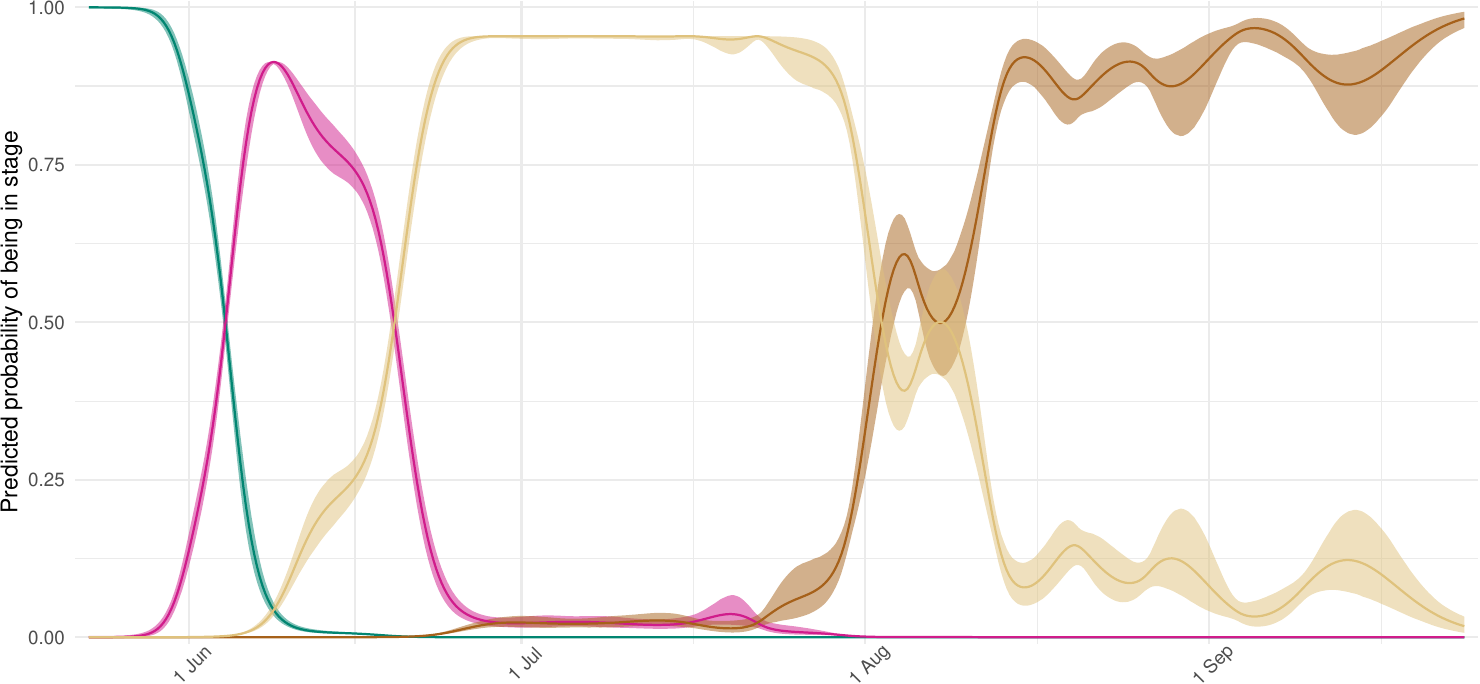}}

}

\caption{\label{fig-plot-nosite}Predicted probabilities of being in a
given phenological stage as a function of time (with 95\% credible
bands) for model \texttt{b1}.}

\end{figure}

Saxifrage is most likely in stage 3, ``senescent'', followed closely by
``open''. \texttt{predict(...,\ type="response")} calculates
\(\mathbb{P}(Z = k \vert \eta) = \mathbb{P}(Z \leq k \vert \eta) - \mathbb{P}(Z \leq k-1 \vert \eta)\).
Figure~\ref{fig-plot-nosite} shows probabilities of being in a stage
given a day of the year using this method (with uncertainty generated
using posterior sampling). It may also be useful to look at
\(\mathbb{P}(Z \leq k \vert \eta) = \sum_{k^\prime=k}^{K} \mathbb{P}(Z = k^\prime \vert \eta)\),
the probability that an observation is in stage \(k\) or lower on a
given day (obtained by summing the \texttt{predict} output
appropriately). Figure~\ref{fig-plot-fit} shows this. The model tracks
the data relatively well for the earlier stages (when sites are more
similar) but that this breaks down as we move to the senescent stage
(and sites become less similar). This indicates how we might proceed to
improve our model.

\subsection{Adding site-specific
effects}\label{adding-site-specific-effects}

\begin{figure}

\centering{

\pandocbounded{\includegraphics[keepaspectratio]{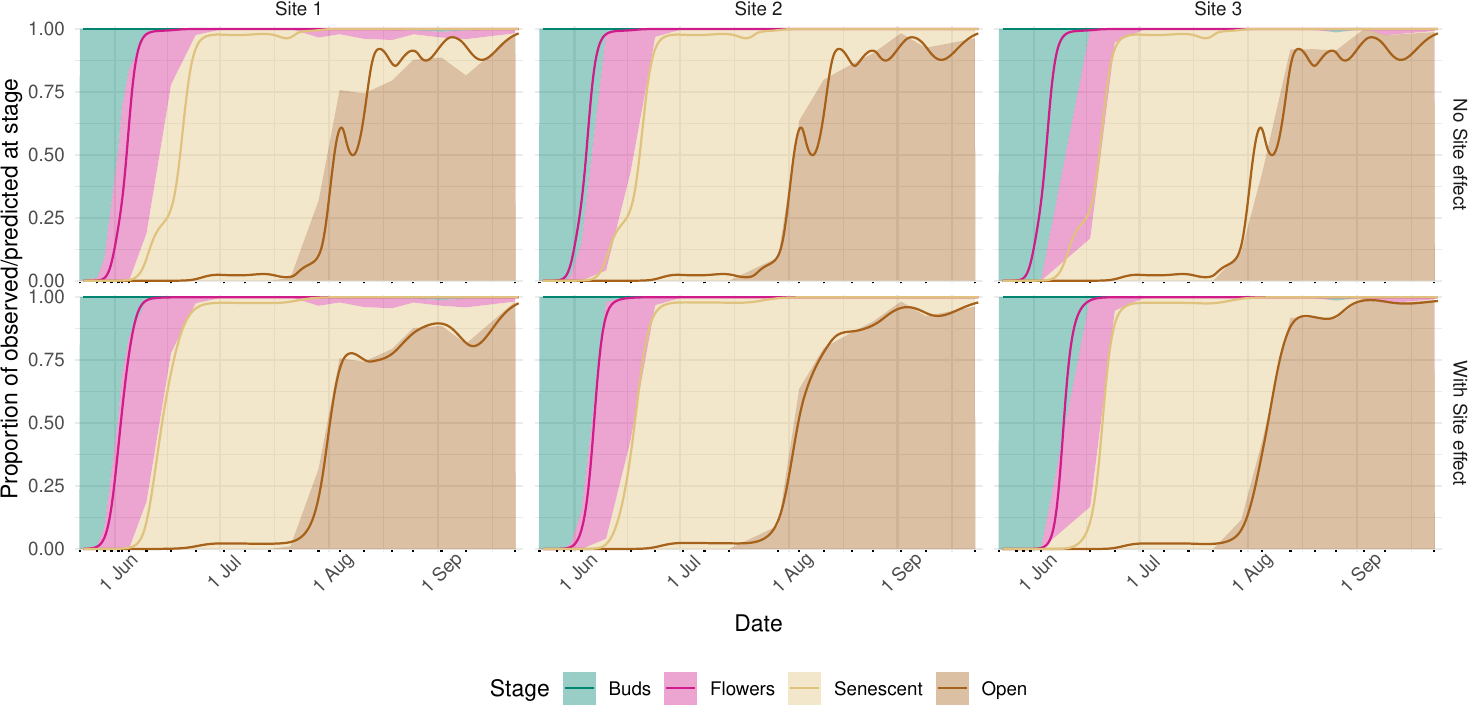}}

}

\caption{\label{fig-plot-fit}Plots showing the relative fit of models
\texttt{b1} (without \texttt{Site} as a covariate; top row) and
\texttt{b2} (when \texttt{Site} was included; bottom row). Background
polygons show the proportion of the data in a given phenological stage
on that date. The lines give the model-based probabilities of being in
that stage or less (top row lines are identical).}

\end{figure}

Although the deviance explained is high for \texttt{b1}, we can see some
fairly large divergences between the data and the model
(Figure~\ref{fig-plot-fit}, top row). The source of these issues may be
that we are ignoring \texttt{Site} in this model, so let's include that
next.

Adding a random effect by site may help explain differences as a change
in intercept for each of the sites (on the linear predictor scale) but
this doesn't necessarily help with the within-year changes we see in the
top row of Figure~\ref{fig-plot-fit}. Instead we model a global temporal
effect with deviations from that effect per site, as necessary. The
simplest and most parsimonious way to do this is via the \texttt{"sz"}
basis in \texttt{mgcv} (see
\texttt{?mgcv::smooth.construct.sz.smooth.spec} for more information and
Pedersen \emph{et al.} (2019) for similar modelling options). As the
interaction between day of year and site treats site as a random effect,
there is shrinkage applied so that each of the per-site smoothers are
only as complex as they need to be.

We now modify the linear predictor in (\ref{eq-model1}) as follows:
\begin{equation}\phantomsection\label{eq-b2}{
\eta_i = \beta_0 + s(\texttt{DOY}_i) + s(\texttt{DOY}_i, \texttt{Site}_i),
}\end{equation} where \(\texttt{Site}_i\) is the site (1, 2, or 3) where
the observation took place.

In R we can write the model as:

\begin{Shaded}
\begin{Highlighting}[]
\NormalTok{b2 }\OtherTok{\textless{}{-}} \FunctionTok{gam}\NormalTok{(iStage }\SpecialCharTok{\textasciitilde{}} \FunctionTok{s}\NormalTok{(doy, }\AttributeTok{k=}\DecValTok{25}\NormalTok{) }\SpecialCharTok{+} \FunctionTok{s}\NormalTok{(doy, Site, }\AttributeTok{bs=}\StringTok{"sz"}\NormalTok{, }\AttributeTok{k=}\DecValTok{25}\NormalTok{),}
          \AttributeTok{data=}\NormalTok{sax\_ocat10, }\AttributeTok{family=}\FunctionTok{ocat}\NormalTok{(}\AttributeTok{R=}\DecValTok{4}\NormalTok{), }\AttributeTok{method=}\StringTok{"REML"}\NormalTok{)}
\end{Highlighting}
\end{Shaded}

The resulting model has an EDF of 14.99 for the day of year ``main''
effect, and 15.06 for the per-site effect. The addition of this
structure has also increased the deviance explained to 91.33\%.
Comparing these models by AIC indicates an almost 1000 point improvement
(\texttt{AIC(b1)=} 8177.25, \texttt{AIC(b2)=} 7263.12).
Figure~\ref{fig-plot-fit} (bottom row) illustrates the improvements in
the fit of the new model, particularly for the ``open'' stage.

\subsection{Residuals}\label{residuals}

\begin{figure}

\centering{

\pandocbounded{\includegraphics[keepaspectratio]{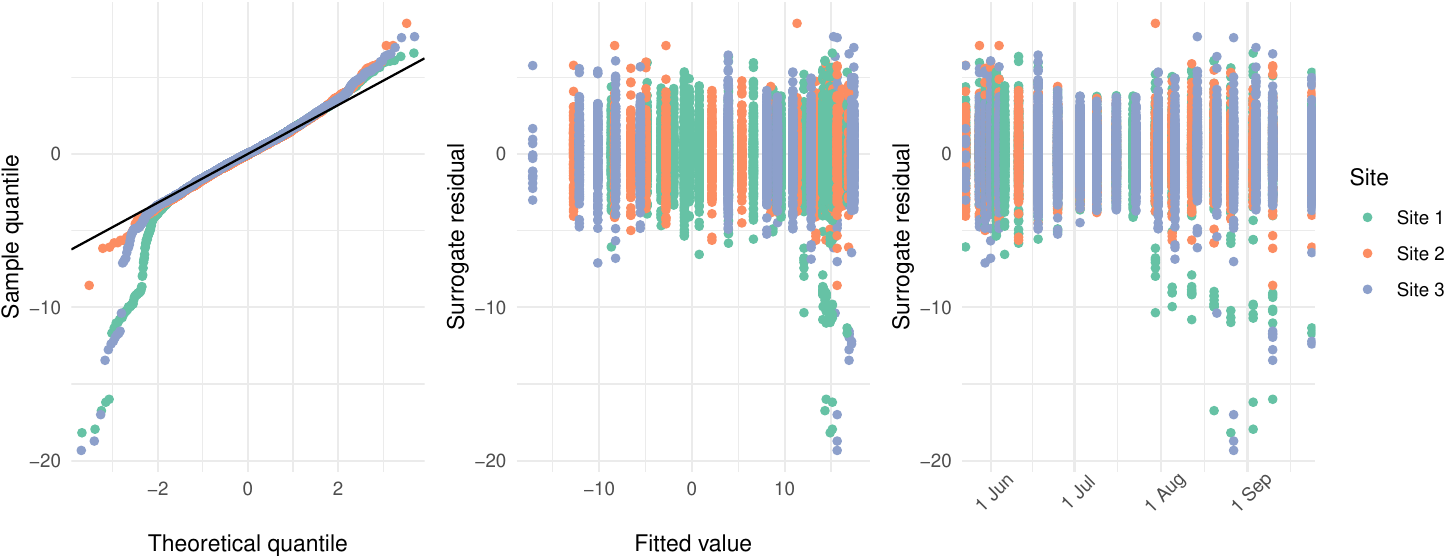}}

}

\caption{\label{fig-sur-resids}Surrogate residual plots for model
\texttt{b2}. Left to right: (\(i\)) quantile-quantile plot showing poor
performance in the lower tails of the distribution; (\(ii\)) residuals
vs.~linear predictor (\ref{eq-b2}), potential evidence of non-constant
variance as the fitted value increases, but investigating further;
(\(iii\)) residuals vs.~day of year shows that larger residuals relate
mostly to the late flowering days Figure~\ref{fig-sax2010-raw}.}

\end{figure}

The \texttt{sure} package (Greenwell \emph{et al.} 2018) provides
\emph{surrogate residuals} (Liu \& and Zhang 2018) which are appropriate
for ordered categorical data. They use a surrogate continuous variable
in place of the discrete response, based on the conditional distribution
of \(\tilde{Z}\) given the observed \(Z\). Surrogate residuals are
randomized, so the random seed must be set for reproducibility.
Figure~\ref{fig-sur-resids} gives useful plots for this data where we
see that poor fit later in the time series is due to secondary flowering
at sites 1 and 3 later in the year (Figure~\ref{fig-sax2010-raw}). These
late flowerings happen while there is both senescence and (majority)
open seeding.

\subsection{Useful plots and
quantities}\label{useful-plots-and-quantities}

The estimated thresholds \(\theta_k\) for \(k=1, \ldots, K\) (ignoring
the \(\pm\infty\) upper/lower bounds) can be obtained via
\texttt{model\$family\$getTheta(TRUE)}. These values give the points
along the scale of the linear predictor (\(\eta\)), given in
(\ref{eq-b2}), that the stages will change. Note since the \(\theta_k\)
must be monotone increasing, they are estimated on a different scale
(McCullagh \& Nelder 1989 sec. 5.1.2), passing the \texttt{TRUE}
argument ensures they are on the \(\theta_k\) scale.

\begin{figure}

\centering{

\pandocbounded{\includegraphics[keepaspectratio]{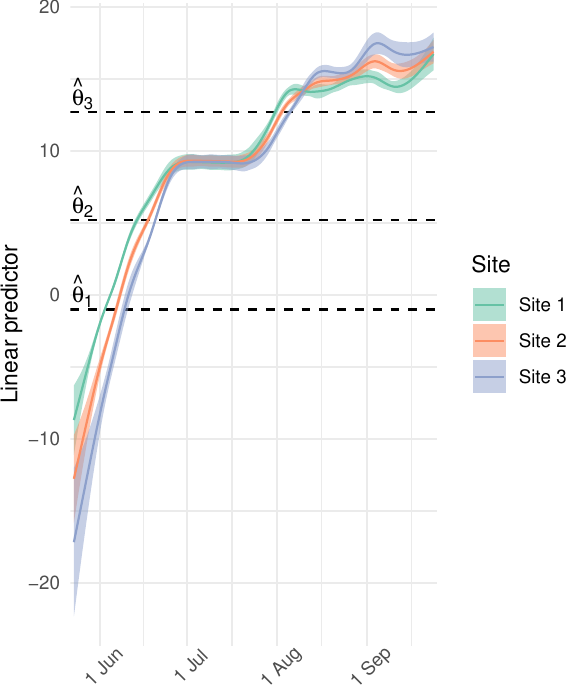}}

}

\caption{\label{fig-cutplot}Linear predictor (logit link scale) values
for model \texttt{b2} for each saxifrage sample site as a function of
date. Dashed lines indicate the thresholds for each phenological stage,
as estimated by the model. When the linear predictor is below
\(\hat{\theta_1}\) plants are budding (according to the model), between
\(\hat{\theta}_1\) and \(\hat{\theta}_2\) they are flowering, and so on.
Bands show 95\% intervals around the linear predictor.}

\end{figure}

Figure~\ref{fig-cutplot} shows the linear predictor for each site, with
the thresholds for reference. Confidence bands around the linear
predictor show the range of dates that the model predicts for that
transition. These were calculated using the method of Nychka (1988), but
could also be calculated using posterior simulation (Miller 2025).

\subsection{Adding year specific effects}\label{yeareff}

To discover more about saxifrage phenology, we might want to look at how
things vary over many years. We now take 10 years of data at site 3 only
and look at how phenology differs between years. An additional
consideration is that sampling only starts once all snow has melted
(\emph{{BioBasis Manual}} 2019). To put our data in the same temporal
frame, we now work with day from first sample (variable \texttt{dffs},
so \texttt{dffs=0} for the first day of sampling in each year; mean melt
day 156, s.d.=14).With this shift, we are now looking at phenology
relative to snow melt for simplicity in this analysis (other temporal
variables may be more appropriate, depending on the goal of the
analysis).

Since this is a much bigger dataset (\(n=\) 31030), we now use
\texttt{mgcv::bam} (Wood \emph{et al.} 2017). \texttt{bam} (``big
additive model'') allows one to fit additive models to very large
datasets more efficiently by parallelizing operations
(\texttt{nthreads=}, to increase the number of CPU cores used to fit the
model), using discretized covariates to reduce memory usage
(\texttt{discrete=TRUE}) and a fast fitting criteria
(\texttt{method="fREML"}). We use the same model formulation as
(\ref{eq-b2}) but use the year as a factor variable (\texttt{fYear}) in
place of \texttt{Site}.

\begin{Shaded}
\begin{Highlighting}[]
\NormalTok{b3 }\OtherTok{\textless{}{-}} \FunctionTok{bam}\NormalTok{(iStage }\SpecialCharTok{\textasciitilde{}} \FunctionTok{s}\NormalTok{(dffs, }\AttributeTok{k=}\DecValTok{25}\NormalTok{) }\SpecialCharTok{+} \FunctionTok{s}\NormalTok{(dffs, fYear, }\AttributeTok{bs=}\StringTok{"sz"}\NormalTok{, }\AttributeTok{k=}\DecValTok{25}\NormalTok{),}
          \AttributeTok{data=}\NormalTok{sax\_ocat1020, }\AttributeTok{family=}\FunctionTok{ocat}\NormalTok{(}\AttributeTok{R=}\DecValTok{4}\NormalTok{),}
          \AttributeTok{method=}\StringTok{"fREML"}\NormalTok{, }\AttributeTok{discrete=}\ConstantTok{TRUE}\NormalTok{, }\AttributeTok{nthreads=}\DecValTok{4}\NormalTok{)}
\end{Highlighting}
\end{Shaded}

We can analyze the results of the \texttt{bam} model exactly as above.
Supplementary Materials A shows plots like those in
Figure~\ref{fig-plot-nosite}, Figure~\ref{fig-plot-fit} and
Figure~\ref{fig-cutplot} for model \texttt{b3}.

To calculate intervals around when we think stage transitions happen
(conditional on the \(\theta_k\) being correct), we can rely on the
assumption that \emph{a posteriori} the model coefficients are
approximately normally distributed with a mean of the estimated
parameters and variance given by the posterior covariance matrix, i.e.,
\(\boldsymbol{\beta} \sim N(\hat{\boldsymbol{\beta}}, \hat{\mathbf{V}}_{\hat{\boldsymbol{\beta}}})\).
We can then sample from that distribution in order to generate possible
\(\boldsymbol{\beta}\)s (under the model), calculate the linear
predictor and find the days where the linear predictor crosses each
\(\theta_k\).

\begin{figure}

\centering{

\pandocbounded{\includegraphics[keepaspectratio]{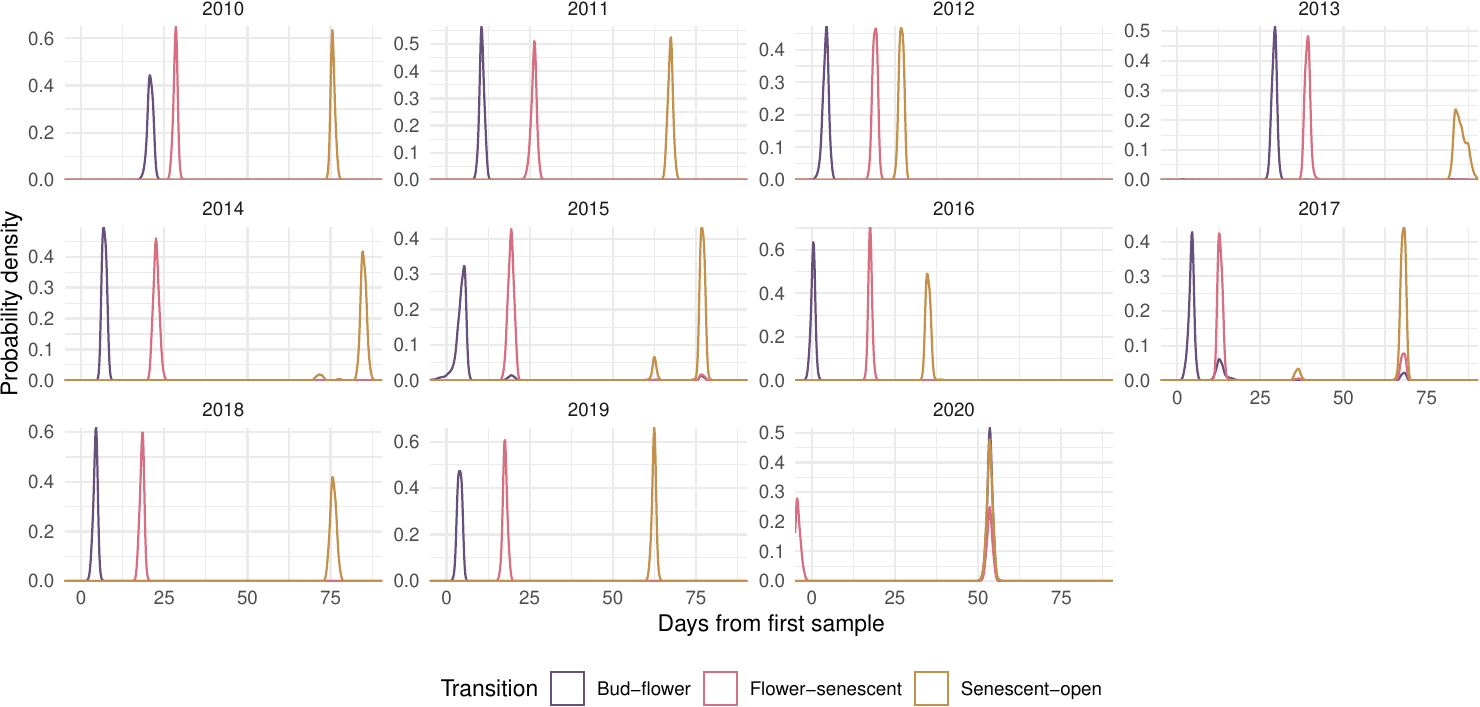}}

}

\caption{\label{fig-transition-density}Probability density of the
transition times for saxifrage in site 3 using model \texttt{b3} where
year is a covariate, giving a sense of uncertainty in transition date.
For each year (facet), we have the kernel density of day of year of
transition from one stage to the next (found using the algorithm given
in the extended paper). Generated assumping that all uncertainty
originates in covariate effects (not the thresholds \(\theta_k\)). Note
that horizontal axes remain the same but vertical axes are on their own
scale per plot.}

\end{figure}%

Results are shown in Figure~\ref{fig-transition-density} as a density
plot over day of year. Saxifrage's potential multi-flowering life
history can be seen as multimodality in the plot, indicating the
multiple flowering times. We can use the data generated for these plots
to obtain uncertainties by simple sample statistics (see extended
paper).

\section{Discussion}\label{disc}

Ignoring the structure in ordinal data, either by treating it as metric
(i.e., where distances are meaningful rather than orderings) or
addressing only pairwise transitions is fraught with issues. Metric
analysis ignores that categories may not be equidistant (transitions
between life stages, say, may not require the same time or energy as
each other; for example, reproduction vs.~hibernation). Liddell \&
Kruschke (2018) highlight additional issues that arise when modelling
ordered categorical data as metric. Modelling only pairwise transitions
will ignore additional uncertainty that we see from e.g., multiple
flowering here.

The GAM framework allows for a wide variety of flexible terms. We can
take into account spatial terms, complex random effects structures,
interactions between variables (via tensor products) and more. See Wood
(2017) for an introduction. \texttt{mgcv} fits an empirical Bayes model,
to move to a fully-Bayesian framework, \texttt{brms} offers almost all
of the flexibility in model terms as \texttt{mgcv} while also allowing
for priors to be specified over the smoothness (variance components) of
the model (Bürkner \& Vuorre 2019). The standard nature of the models
addressed here (and the ability to generate from the posterior) allows
us to explore other inferential outputs, such as rates of change and
number of days the organism is in a particular stage.

Further technical detail on these models can be found in Fahrmeir \&
Tutz (2001) (Section 3.3.1), Fahrmeir \emph{et al.} (2013) (Section
6.3.1) and Wood \emph{et al.} (2016) (Appendix K). It may be necessary
to fit more complicated models when data have additional or differing
structure. These topics have been explored in such publications as:
Chambert \emph{et al.} (2015), Hufkens \emph{et al.} (2018), Bürkner \&
Vuorre (2019) and Boersch-Supan \emph{et al.} (2024).

The saxifrage data are not longitudinal in nature: individual plants are
not tracked through time, instead we are making inference at the
population-level. We can fit longitudinal models via the addition of
individual covariates and random effects if we do identify plants. Note
that the model cannot include covariates that vary by stage, as these
can lead to negative probabilities (see Bürkner \& Vuorre (2019),
Appendix A).

Additional quantities of interest can also be derived. Times at which
given proportion of the samples have transitioned to a given state
(e.g., the day on which at least 50\% of the population are in the open
stage, or later) can be obtained. Formally, we want to find the value of
\(m\) such that \(\mathbb{P}(Z \geq k \vert \eta, \text{DOY}=m) = p\) if
\(p\) is the proportion of the population we are interested in. This can
be found via a simple grid search over the days of the year in our case
(it may require an optimization in other cases). The analyses presented
here show how the proportions of a population in different stages change
with time. Taking derivatives allows us to look at changes in the
instantaneous rate of change to be derived, which can lead to further
insights (since the derivative of the cumulative distribution function
is the probability density function).

Phenology is just one area of ecology where we might find ordered
categorical data. There are many situations in which we record
categorical data which has an ordering but categories are not guaranteed
to be equidistant. This article has shown some particular
application-specific plots and ideas for those interested in phenology,
but the machinery involved (\texttt{mgcv}) is widely-applicable.

\section{Acknowledgements}\label{acknowledgements}

I would like to thank Jeremy Greenwood (University of St Andrews) and
Bruce Lynch for drawing my attention to phenological data. Catriona
Morrison provided many very useful comments on an early version of the
paper. Massive thanks go to Simon Wood for his continued work improving
the \texttt{mgcv} package. Data from the Greenland Ecosystem Monitoring
Programme were provided by the Department of Bioscience, Aarhus
University, Denmark in collaboration with Greenland Institute of Natural
Resources, Nuuk, Greenland, and Department of Biology, University of
Copenhagen, Denmark. Data are available at
\url{https://doi.org/10.17897/YXH1-ZB25}.

\section*{References}\label{references}
\addcontentsline{toc}{section}{References}

\phantomsection\label{refs}

\end{document}